# Ultraprecise optical-frequency stabilization with heterogeneous III-V/Si lasers

Liron Stern[1,2,5], Wei Zhang[1,2,6], Lin Chang[3], Joel Guo[3], Chao Xiang[3], Minh A. Tran[3,7], Duanni Huang[3], Jonathan D. Peters[3], David Kinghorn[3,4], John E. Bowers[3], and Scott B. Papp[1,2,*]

[1]National Institute of Standards and Technology, Time and frequency division, 325 Broadway, Boulder, CO 80305, USA
[2]Department of Physics, University of Colorado, Boulder, Colorado 80305, USA
[3]Electrical and Computer Engineering Department, University of California Santa Barbara, California 93106, USA
[4]Pro Precision Process & Reliability, Santa Barbara, CA 93106, USA
[5]Current address: Department of Applied Physics, Hebrew University of Jerusalem, Israel
[6]Current address: Jet Propulsion Laboratory, 4800 Oak Grove Drive Pasadena, CA 91109
[7]Current address: Nexus Photonics, 6500 Hollister Ave., Ste 140, Goleta, CA 93117, USA
*Corresponding author: scott.papp@nist.gov

**Demand for low-noise, continuous-wave, frequency-tunable lasers based on semiconductor integrated photonics has been advancing in support of numerous applications. In particular, an important goal is to achieve narrow spectral linewidth, commensurate with bulk-optic or fiber-optic laser platforms. Here, we report on laser-frequency-stabilization experiments with a heterogeneously integrated III/V-Si widely tunable laser and a high-finesse, thermal-noise-limited photonic resonator. This hybrid architecture offers a chip-scale optical-frequency reference with an integrated linewidth of 60 Hz and a fractional frequency stability of $2.5 \times 10^{-13}$ at 1-second integration time. We explore the potential for stabilization with respect to a resonator with lower thermal noise by characterizing laser-noise contributions such as residual amplitude modulation and photodetection noise. Widely tunable, compact and integrated, cost effective, stable and narrow linewidth lasers are envisioned for use in various fields, including communication, spectroscopy, and metrology.**

The linewidth and frequency jitter of lasers, and its corresponding phase-noise power-spectral density, play a significant role in innovation among a variety of application areas that benefit from miniature or scalable technology, including optical-atomic clocks, photonic microwave signal generation and analysis, sensing with light, and optical communication [1–5]. Tabletop-scale lasers based on sophisticated design with bulk components are widely employed, but lower cost, scalable integrated-photonics laser technologies can open up new opportunities, especially using intricate design of nanophotonics to control radiation.

The highest performance tabletop-scale narrow-linewidth lasers utilize a hybrid approach in which the laser itself is optimized for low optical-frequency noise at high Fourier frequency and a separate optical cavity is optimized for long-term frequency stability and insensitivity to environmental fluctuations. Such stable laser systems usually involve meticulous design of laser and optical systems and their supporting enclosures and electronics [6–8]. Hence, it is important to realize narrow Lorentzian linewidth with the laser as well as excellent intrinsic properties for laser stabilization such as low intensity noise, residual amplitude modulation, and wideband frequency control.

Recently, there has been significant effort to introduce stable lasers aiming for field applications in which size, weight, and power consumption (SWAP), robustness and integration are primary considerations. To that end, two significant core elements of stable laser systems, namely the laser and the cavity, are continuously being improved both in performance and SWAP. Indeed, a significant effort towards reducing the core dimensions of spectrally pure integrated lasers has been demonstrated, using several approaches such as Brillouin lasers [9–12], hybrid semiconductor lasers [13–15] and heterogeneous semiconductor lasers [16,17]. Simultaneously, reduction in the dimensions of optical-cavity systems has also been pursued, using evacuated ultra-low expansion cavities [18–21], self-injection locking [22,23], and whispering gallery mode resonators [24–32].

In this paper, we demonstrate and explore a compact, narrow linewidth laser system by combining a widely tunable, low-noise III/V-Si heterogeneously integrated laser with a monolithic, compact, high-finesse photonic resonator. The integrated laser is composed of a III/V gain section, and Si waveguides and resonators. These integrated laser [33] and photonic resonator [34] components have been developed recently by the authors. Both devices have been packaged for robust use in the lab, including electrical and polarization-maintaining optical fiber connections for the integrated laser, and a thermal- and vibration-mitigation holding structure for the photonic resonator. By hybridizing these technologies, which have nominally been envisioned separately for data communications applications and optical-frequency metrology, we demonstrate a laser system with an integrated linewidth of ≈60 Hz and a frequency noise floor of 22 Hz$^2$/Hz. The fractional frequency instability of the laser reaches a minimum of $2.5 \times 10^{-13}$ at 1 sec integration time. Furthermore, we assess the laser-frequency noise contributions from the systems' intrinsic properties such as residual amplitude modulation and intensity noise.

Figure 1 presents a diagram of the integrated laser (Fig. 1a) and its wavelength-tuning capability (Fig. 1b), and a photograph of the cylindrical, fused-silica photonic resonator (Fig. 1c). We briefly recount the techniques to realize these components; see Refs. [5,33,34] for more details. The widely tunable, narrow-linewidth laser is based on a heterogeneous silicon photonics platform with the following elements: a Sagnac loop front mirror, a gain section realized by bonding and processing III-V material on top of a pre-patterned silicon rib waveguide, a phase control section, and a back mirror reflector formed by dual-ring silicon resonators. The phase sections and rings are thermo-optically tuned with resistive heaters. The multi-ring, resonator-based mirror simultaneously provides wide wavelength tunability through the Vernier effect and linewidth reduction due to the effective cavity length enhancement at resonance. Such designs enable Lorentzian linewidth in the range of 100–1000 Hz. Yet low Fourier frequency fluctuations of the laser output frequency makes frequency stabilization with respect to an optical-reference cavity appealing, when necessary for applications like sensing and spectroscopy. The integrated laser provides wavelength tuning range of 40-120 nm centered at the telecom bands. For the dual-ring resonator mirror design used, typical tuning range is over 60 nm with over 50 dB side mode suppression ratio (SMSR) across the entire range, illustrated in Fig. 1b. Additionally, we include an on-chip semiconductor optical amplifier to provide higher power than the laser oscillator alone, highlighting the capability for design with heterogeneously integrated photonics. The gain sections and silicon passives are all fully integrated on SOI and are all electrically tunable. As opposed to hybrid lasers consisting of two chips aligned to each other, such lasers offer superior resistance to shock as well as packaging simplicity.

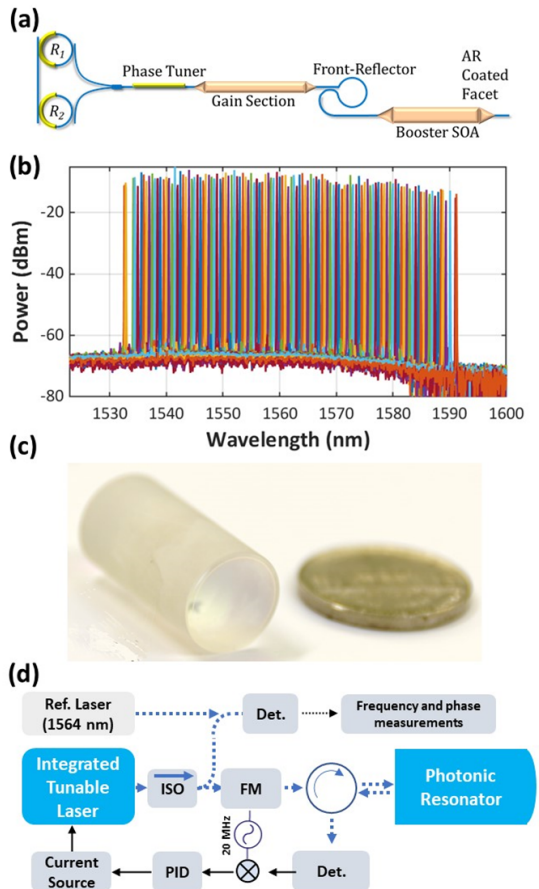

Fig. 1. (a) Schematic of integrated laser. (b) Optical spectra of the integrated laser, supporting 60 nm of wavelength tuning. (c) Photograph of a fused-silica photonic resonator. (d) Stabilization of the integrated laser and frequency metrology. FM – phase modulator, ISO – optical isolator, DET – photodetector.

The photonic resonator is based on a bulk, fused-silica plano-convex cylinder with a length of 25.4 mm, a diameter of 12.7 mm, and a radius of curvature of 0.5 m. A best choice of fused-silica material enables a high-finesse photonic resonator by way of low scattering and absorption losses in the material, low scattering super-polished facets, and low scattering and absorption reflection coatings. In the present photonic resonators, we have realized a finesse of ≈20,000. The ion-beam-sputtered reflection coating for the dual wavelength ranges of 1550 nm and 780 nm is measured to be ≈0.99998, whereas the scattering and loss due to the silica material is measured to be 130 part-per-million (ppm). Hence, the cavity offers a Lorentzian linewidth of ≈200 kHz, corresponding to a finesse of ≈20,000, and quality factor of ≈1 billion. Our monolithic photonic resonator, without the evacuated center bore essential in Fabry-Perot cavity, eliminates the need for high vacuum and composite materials. We design the photonic resonator holding mount to provide thermal and vibration isolation from the environment, using the favorable material properties of Teflon and an aluminum enclosure as reported in [34]. By stabilizing a commercial laser, which is not amendable to integration and scalable manufacturing, to the photonic resonator, we have previously identified the ~25 Hz thermal-noise-limited integrated linewidth that is achievable [34].

We perform optical-frequency stabilization of the integrated laser with the photonic resonator, using the Pound-Drever-Hall scheme; see Fig. 1d for a schematic of the experimental apparatus. Here the integrated laser, operating at the wavelength of 1564 nm, is powered by a low-noise current source and temperature-controlled with a thermo-electric cooler. Wavelength tuning of the integrated laser is achieved by actuating three integrated heaters that control a phase section and two ring resonators. After the fiber coupled laser output is brought off-chip, we utilize a fiber isolator and a fiber coupler to create two portions of the light. The first is to characterize the laser frequency fluctuations with respect to a table-top, ~1 Hz linewidth, ultra-low-expansion, cavity-stabilized laser, also operating at the wavelength of 1564 nm. In order to derive a Pound-Drever-Hall (PDH) error signal, the second portion of the laser (~10 μW) is phase modulated by a standard, fiber-coupled lithium-niobate waveguide modulator and mode matched to excite a single transverse mode of the photonic resonator. An optical circulator assists in detecting back reflected light, which is demodulated and sent to a proportional integral differential (PID) controller to regulate the current fed to the laser.

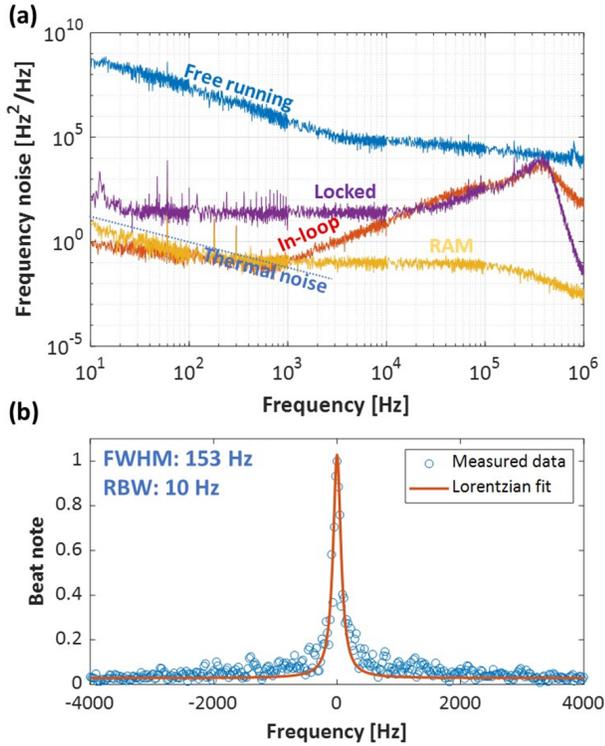

Fig. 2. (a) Frequency-noise PSD: Free-running integrated laser (blue trace), stabilized integrated laser (purple trace), PDH in-loop error (red trace), residual amplitude modulation (RAM, yellow trace), and the thermorefractive thermal noise of the photonic-resonator (blue line). (b) Instantaneous lineshape (blue points) of the stabilized integrated laser, which is measured by heterodyne with the reference laser and with 10 Hz resolution bandwidth (RBW). A fit to the data yields the linewidth of the laser.

We characterize the linewidth of the stabilized integrated laser with several different, but complementary techniques. We measure the integrated laser frequency-noise power spectral density (PSD) by conversion of the optical heterodyne beatnote obtained with the reference laser to a voltage and subsequent realtime digitization and Fourier analysis. Figure 2a presents the frequency-noise spectrum of the free running (blue) and stabilized (purple) integrated laser, and the thermal noise of the photonic resonator (blue line). To obtain the locked data, we use an electronic low-pass filter at ~300 kHz on the heterodyne signal to improve the dynamic range of our measurement system. Clearly, locking the laser to the photonic resonator significantly reduces frequency noise below a Fourier frequency of ~1 MHz, within the bandwidth of the PID loop. For instance, the frequency-noise power spectral density of the cavity-stabilized laser is 130 $Hz^2Hz^{-1}$ at 10 Hz Fourier frequency, corresponding to ≈4x$10^6$ reduction of noise of the free-running laser.

Frequency-noise analysis provides further details of system operation, particularly intrinsic contributions to the stabilized laser frequency from the hybrid system. In Fig. 2a, we present measurements of residual-amplitude modulation (RAM, yellow trace) and in-loop error (red trace) converted to frequency-noise power spectra density. RAM arises from an imbalance of the phase modulation sidebands that we characterize by the PDH baseline off-resonance [7,32,34]. In-loop error characterizes an inability to resolve or correct frequency noise. The RAM and in-loop noise coincide at low Fourier frequency, and they are below the stabilized laser noise. Beyond 10 kHz, the frequency noise is dominated by the in-loop error of the 350 kHz bandwidth PDH lock.

We determine the total laser linewidth of 60 Hz by integrating phase noise from 1 MHz to the Fourier frequency corresponding to 1 $rad^2$. A similar analysis, using the free running laser frequency noise leads to an integrated linewidth of ≈65 kHz. The dynamic range of our frequency-noise PSD measurement system does not adequately resolve the stabilized laser noise from ~10 Hz to 10 kHz, due to the free-running frequency-noise floor of the laser at ~1 MHz. However, a reduction in the Lorentzian linewidth by a factor of two with lower laser-cavity losses or an improvement in the PDH servo bandwidth, or use of a wider dynamic range digitizer would alleviate this issue. Indeed, in terms of the potential for Hz-level or better frequency stabilization of the integrated laser, the RAM and in-loop contributions confirm the possibility for a <10 Hz linewidth, given the level of these contributions at the lowest Fourier frequency of 10 Hz that we have been able to measure.

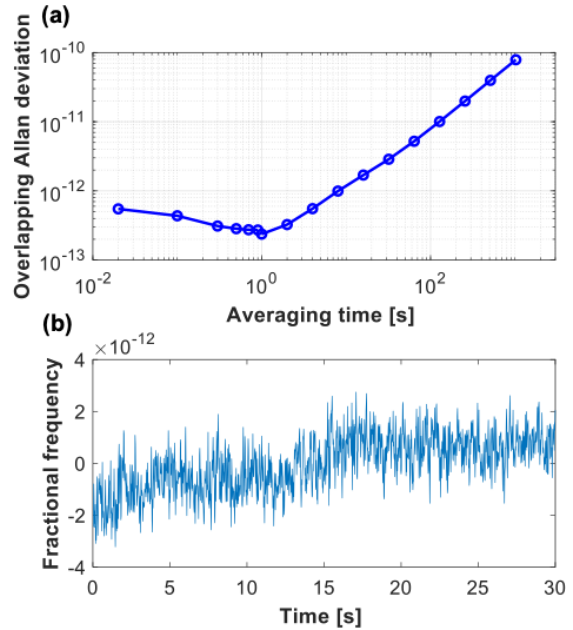

Fig. 3. (a) Overlapping Allan deviation of the integrated laser and photonic resonator system, and (b) Measured fractional frequency of the stabilized integrated laser via the heterodyne beat signal as a function of time.

A second laser linewidth estimate is based directly on the optical heterodyne with the reference laser. Figure 2b shows a typical instantaneous optical lineshape of the stabilized laser. The full-width-half-maximum (FWHM) of the laser's optical lineshape is ≈150 Hz. Obtaining a quantitative linewidth estimate from this data is challenging due to the observed <100 Hz/s drift of the photonic resonator.

We obtain additional statistical information about the stabilized integrated laser by time-domain analysis of the optical heterodyne with the reference laser, which has a frequency drift below 1 Hz/s. Figure 3a presents the fractional-frequency stability obtained from an overlapping Allan deviation computation of the heterodyne signal, recorded with a dead-time-free electronic counter. This calculation does not remove frequency drift. The stabilized integrated laser reaches an Allan deviation of $2.5 \times 10^{-13}$ at $\tau$ = 1 s, and it follows $\sim 1 \times 10^{-13} \tau$ at longer times as a consequence of the thermal drift of the photonic resonator. Figure 3b presents a 30 second portion of the heterodyne frequency record, indicating our observed typical behavior of the stabilized laser frequency output.

In summary, we have demonstrated a compact, narrow linewidth laser system by combining a widely tunable, low-noise III/V-Si heterogeneously integrated laser with a monolithic, compact, high-finesse photonic resonator. The instability of the laser reaches a minimum fractional frequency of $2.5 \times 10^{-13}$ at 1-sec integration time. Moreover, the Lorentzian linewidth of the stabilized integrated laser is at the kHz level, providing the capability for high signal-to-noise measurements, optical and microwave signal generation, and data encoding. As our laser is tunable across the entire C-band and the resonator supports multiple high-finesse modes at a bandwidth of tens of nm, we expect narrow linewidth performance across for example the entire C band. Towards improved long-term stability we envision a subsequent step of stabilization to atomic transitions at the cost of wavelength tuning [34]. The integrated laser and photonic resonator technologies in this work are both commercially available, opening opportunities for wider use of the technology in view of the practical results and limitations presented here.

We thank Esther Baumann for help and Jizhao Zang and Jennifer Black for reading the paper. This work was funded by Defense Advanced Research Projects Agency (DARPA) under DODOS (HR0011-15-C-055) and A-PhI (FA9453-19-C-0029) programs, and NIST.

**References.**
[1] T. C. Briles, J. R. Stone, T. E. Drake, D. T. Spencer, C. Fredrick, Q. Li, D. Westly, B. R. Ilic, K. Srinivasan, S. A. Diddams, and S. B. Papp, Opt. Lett., OL **43**, 2933 (2018).
[2] T. E. Drake, T. C. Briles, J. R. Stone, D. T. Spencer, D. R. Carlson, D. D. Hickstein, Q. Li, D. Westly, K. Srinivasan, S. A. Diddams, and S. B. Papp, Phys. Rev. X **9**, 031023 (2019).
[3] Y.-H. Lai, Y.-K. Lu, M.-G. Suh, Z. Yuan, and K. Vahala, Nature **576**, 65 (2019).
[4] A. Beling, X. Xie, and J. C. Campbell, Optica **3**, 328 (2016).
[5] M. A. Tran, D. Huang, and J. E. Bowers, APL Photonics **4**, 111101 (2019).
[6] M. Notcutt, L.-S. Ma, J. Ye, and J. L. Hall, Opt. Lett., OL **30**, 1815 (2005).
[7] J. M. Robinson, E. Oelker, W. R. Milner, W. Zhang, T. Legero, D. G. Matei, F. Riehle, U. Sterr, and J. Ye, Optica, OPTICA **6**, 240 (2019).
[8] S. Häfner, S. Falke, C. Grebing, S. Vogt, T. Legero, M. Merimaa, C. Lisdat, and U. Sterr, Opt. Lett., OL **40**, 2112 (2015).
[9] I. S. Grudinin, A. B. Matsko, and L. Maleki, Phys. Rev. Lett. **102**, 043902 (2009).
[10] H. Lee, T. Chen, J. Li, K. Y. Yang, S. Jeon, O. Painter, and K. J. Vahala, Nat Photon **6**, 369 (2012).
[11] W. Loh, J. Becker, D. C. Cole, A. Coillet, F. N. Baynes, S. B. Papp, and S. A. Diddams, New Journal of Physics **18**, 045001 (2016).
[12] S. Gundavarapu, G. M. Brodnik, M. Puckett, T. Huffman, D. Bose, R. Behunin, J. Wu, T. Qiu, C. Pinho, N. Chauhan, J. Nohava, P. T. Rakich, K. D. Nelson, M. Salit, and D. J. Blumenthal, Nature Photonics **13**, 60 (2019).
[13] B. Stern, X. Ji, A. Dutt, and M. Lipson, Opt. Lett., OL **42**, 4541 (2017).
[14] H. Guan, A. Novack, T. Galfsky, Y. Ma, S. Fathololoumi, A. Horth, T. N. Huynh, J. Roman, R. Shi, M. Caverley, Y. Liu, T. Baehr-Jones, K. Bergman, and M. Hochberg, Opt. Express, OE **26**, 7920 (2018).
[15] Y. Zhu and L. Zhu, Opt. Express, OE **27**, 2354 (2019).
[16] C. T. Santis, S. T. Steger, Y. Vilenchik, A. Vasilyev, and A. Yariv, PNAS **111**, 2879 (2014).
[17] D. Huang, M. A. Tran, J. Guo, J. Peters, T. Komljenovic, A. Malik, P. A. Morton, and J. E. Bowers, Optica, OPTICA **6**, 745 (2019).
[18] D. R. Leibrandt, M. J. Thorpe, M. Notcutt, R. E. Drullinger, T. Rosenband, and J. C. Bergquist, Opt. Express, OE **19**, 3471 (2011).
[19] B. Argence, E. Prevost, T. Lévèque, R. L. Goff, S. Bize, P. Lemonde, and G. Santarelli, Opt. Express, OE **20**, 25409 (2012).
[20] J. Davila-Rodriguez, F. N. Baynes, A. D. Ludlow, T. M. Fortier, H. Leopardi, S. A. Diddams, and F. Quinlan, Opt. Lett., OL **42**, 1277 (2017).
[21] D. Świerad, S. Häfner, S. Vogt, B. Venon, D. Holleville, S. Bize, A. Kulosa, S. Bode, Y. Singh, K. Bongs, E. M. Rasel, J. Lodewyck, R. Le Targat, C. Lisdat, and U. Sterr, Scientific Reports **6**, 33973 (2016).
[22] W. Liang, V. S. Ilchenko, D. Eliyahu, A. A. Savchenkov, A. B. Matsko, D. Seidel, and L. Maleki, Nature Communications **6**, 7371 (2015).
[23] H. Patrick and C. E. Wieman, Review of Scientific Instruments **62**, 2593 (1991).
[24] A. B. Matsko, A. A. Savchenkov, N. Yu, and and Lute Maleki, J. Opt. Soc. Am. B **24**, 1324 (2007).
[25] J. Alnis, A. Schliesser, C. Y. Wang, J. Hofer, T. J. Kippenberg, and T. W. Hänsch, Phys. Rev. A **84**, 011804 (2011).
[26] I. Fescenko, J. Alnis, A. Schliesser, C. Y. Wang, T. J. Kippenberg, and T. W. H nsch, Opt. Express **20**, 19185 (2012).

[27] L. M. Baumgartel, R. J. Thompson, and N. Yu, Opt. Express, OE **20**, 29798 (2012).
[28] H. Lee, M.-G. Suh, T. Chen, J. Li, S. A. Diddams, and K. J. Vahala, Nature Communications **4**, 2468 (2013).
[29] W. Weng, J. D. Anstie, T. M. Stace, G. Campbell, F. N. Baynes, and A. N. Luiten, Phys. Rev. Lett. **112**, 160801 (2014).
[30] W. Loh, A. A. S. Green, F. N. Baynes, D. C. Cole, F. J. Quinlan, H. Lee, K. J. Vahala, S. B. Papp, and S. A. Diddams, Optica **2**, 225 (2015).
[31] J. Lim, A. A. Savchenkov, E. Dale, W. Liang, D. Eliyahu, V. Ilchenko, A. B. Matsko, L. Maleki, and C. W. Wong, Nature Communications **8**, 8 (2017).
[32] W. Zhang, F. Baynes, S. A. Diddams, and S. B. Papp, Phys. Rev. Applied **12**, 024010 (2019).
[33] M. A. Tran, D. Huang, J. Guo, T. Komljenovic, P. A. Morton, and J. E. Bowers, IEEE Journal of Selected Topics in Quantum Electronics **26**, 1 (2020).
[34] W. Zhang, L. Stern, D. Carlson, D. Bopp, Z. Newman, S. Kang, J. Kitching, and S. B. Papp, Laser & Photonics Reviews **14**, 1900293 (2020).